\documentclass[11pt,showpacs]{revtex4}
\input epsf.sty
\usepackage{sidecap}
\usepackage{ulem}
\usepackage{cancel}
\usepackage{color}

\def\rr#1{\textcolor{red}{#1}}

\def\comment#1{}

scaled1000  
scaled600

\usepackage{graphicx}
\begin{document}

\title{High energy neutrino oscillation at the presence of the Lorentz Invariance Violation }

\author{\bf Iman Motie$^{(a)}$ and \bf She-Sheng Xue$^{(b)}$}
\affiliation{$^{(a)}$Department of Physics, Isfahan University of
Technology, Isfahan 84156-83111, Iran}
\email{i.motie@ph.iut.ac.ir}
\affiliation{$^{(b)}$ICRANeT Piazzale della Repubblica, 10 -65122, Pescara,\\
Dipartimento di Fisica, University of Rome ``La Sapienza", Rome,
Italy}

\email{xue@icra.it}

\date{Received version \today}

\begin{abstract}
Due to quantum gravity fluctuations at the Planck scale, the
space-time manifold is no longer continuous, but discretized. As a
result the Lorentz symmetry is broken at very high energies. In
this article,  we study the neutrino oscillation pattern due to
the Lorentz Invariance Violation (LIV), and compare it with  the
normal neutrino oscillation pattern due to neutrino masses. We
find that at very high energies,  neutrino oscillation pattern is
very different from the normal one. This could provide an
possibility to study  the Lorentz Invariance Violation by
measuring the oscillation pattern of very high energy neutrinos
from a cosmological distance.
\end{abstract}

 \maketitle

\section{Introduction}\label{intro}
Nowadays the violation of the Lorentz symmetry  has attracted
increasing attention, because the Lorentz symmetry is the main
concept of special relativity and any relativistic theory which is
invariant under continuous Lorentz transformations. A growing
number of speculations suggests that Lorentz invariance might be
violated or deformed at very high energies
\cite{tati1972}\cite{lv2}. The local Lorentz symmetry has been
examined in many sectors of the standard model (SM) relating to
photons, electrons, protons, and neutrons \cite{exp1}-\cite{exp3},
and none of Lorentz invariance violation (LIV) has been identified
so far in these sectors for low-energies. The Lorentz invariance
should be violated at very high energy scale or the Planck scale,
since the Lorentz group is unbounded at the high boost (or high
energy) end, in principle it might subject to modifications in the
high boost limit \cite{coleman,ste}. The Lorentz symmetry is based
on the assumption that space-time is scale-free, namely there is
no fundamental length scale associated with the Lorentz group.
However, due to violent fluctuations of quantum gravity at the
Planck scale $M_{\rm pl}= \hbar c/\lambda_{\rm pl}\sim 10^{19}
$GeV, $\lambda_{\rm pl} = \sqrt{\hbar G/c^3}=10^{-33}$cm, the
space-time manifold is no longer continuous, but discretized, and
as a consequence, the Lorentz symmetry is broken.  The
discretization (foam structure) of space-time manifold with a
minimal spacing $\sim\lambda_{\rm pl}$ was first discussed by
Wheeler \cite{wheeler1964}, and have been intensively studied in
literatures (see for example \cite{preparata91,xue2010}). In
Ref.~\cite{anb} , by using the universal entropy bound, it has
been shown that the space-time has a minimum length scale
proportional to the Planck length, leading to a discrete
space-time structure.

The possibilities of Lorentz invariance violation have been
studied in quantum-gravity models \cite{qg}, string theory
\cite{string1,string2}, Loop gravity \cite{lqg1,lqg2},
non-commutative geometry \cite{ncom1}-\cite{ncom3},the doubly
special relativity (DSR)\cite{dsr}. In addition, there are some
other effective field theories for Lorentz violation, for
examples, the Coleman- Glashow model \cite{coll}, the minimal
standard model extension (SME) \cite{kas}, and the newly proposed
standard model supplement (SMS)\cite{sms1,sms2}.

In recent years, there has been much interest in testing LIV
effects. However, observational tests face a major obstacle of
practical nature: LIV effects due to quantum-gravity are expected
to be extremely small because of Planck-scale suppression, and
low-energy measurements are likely to require very high
sensitivities \cite{lv-review}. This leads to the use of high
energy astrophysics data to provide constraints on LIV effects.
For examples, Gamma-Ray Burst (GRB) data are analyzed to see LIV
effects on the arrival time of photons at different energies
\cite{grb}. However, high energy photons can be annihilated via
pair creation with the IR background, and this limits the
distances that high energy photons can travel, and the photon
number fluxes lower for higher energies limit \cite{hphoton}.

Very high energy neutrinos \cite{hi_nu} provide an alternative to
test LIV effects. Practically all current GRB models \cite{ngrb}
predict bursts of very high energy neutrinos, with energy ranging
from 100 TeV to $10^4$ TeV (and possibly up to $10^6$ TeV)
\cite{13piran}\cite{14piran}. In addition, neutrinos with energy
up to $\sim 10^{21}$ eV are supposed to be produced by
cosmological objects like GRB and Active Galactic Nuclei (AGN)
\cite{proth}. These high-energy neutrinos from cosmological
distances can open a new window on testing LIV effects. It was
suggested that neutrinos of energies as high as $10^{22} -
10^{24}$ eV could be produced by topological defects like cosmic
strings, necklaces and domain walls \cite{wall}. Theoretical
framework for Lorentz violation and neutrino oscillation
probabilities is proposed in Ref.~\cite{nbq}.

In this article, we formulate the discretization of space-time
manifold as a hyperbolic lattice with the lattice spacing $a$, and
adopt the Lorentz-symmetry-breaking Wilson operator \cite{wilson1}
for neutrino fields on the lattice. We show that neutrino
oscillations depend not only on their non-vanishing masses, but
also on the Lorentz-symmetry-breaking Wilson term in high
energies. This provides the possibility to test LIV effects by
studying high-energy cosmic neutrinos oscillations.

\section{Bosons and fermions on a discrete space-time}\label{kogut}

We first give a brief review on the energy-momentum relation of
free bosons and fermions on a hypercubic lattice of space-time.
The Klein-Gordon equation for a free boson field $\phi(\vec{x},t)$
in $3+1$ dimension space-time,
\begin{eqnarray}
\ddot{\phi}=\nabla^2\phi-m^2\phi,
\label{kg}
\end{eqnarray}
where $m$ is the boson mass. Eq.~(\ref{kg}) gives the
energy-momentum relation of the free boson field
\begin{eqnarray}
E^2=k^2+m^2. \label{dr}
\end{eqnarray}
In a hypercubic spatial lattice (time continuum), one can write \cite{kogut}
\begin{eqnarray}
a^2\nabla^2\phi\rightarrow \phi (\mathbf{n}+a)+\phi
(\mathbf{n}-a)-2\phi (\mathbf{n})=(\mathbf{d}^{\,+}
+\mathbf{d}^{\,-} -2)\phi (\mathbf{n}), \label{dif1}
\end{eqnarray}
where 3-dimension vector $\mathbf{n} \equiv a\,(n_1, n_2, n_3)$,
$a$ is the lattice spacing and the shift operators
$\mathbf{d}^{\,\pm}\phi(\mathbf{n})=\phi(\mathbf{n}\pm a)$. Then,
the Klein-Gordon equation on the lattice is given by
\begin{eqnarray}
\ddot{\phi}=(\mathbf{d}^{\,+} +\mathbf{d}^{\,-} -2)\phi
(\mathbf{n})-m^2\phi, \label{kg2}
\end{eqnarray}
and the energy-momentum relation on the lattice is given by
\begin{eqnarray}
E^2=m^2-\frac{2[\cos({k}\cdot a)-1]}{a^2}, \label{kg3}
\end{eqnarray}
Eqs.~(\ref{dif1}-\ref{kg3}) are not invariant under the Lorentz
transformations. For low-energy particles $k\cdot a\ll 1$, the
energy-momentum relation can approximately be written by
\begin{eqnarray}
E^2=m^2+k^{\,2}-\frac{1}{12}(k^4 a^2)+\mathcal O(k^6 a^4),
\label{kg4}
\end{eqnarray}
which approaches the energy-momentum relation (\ref{dr}) and the Lorentz symmetry is restored.
Assuming a more complicate discretization of space-time, we parameterize the energy-momentum relation as
\begin{eqnarray}
E^2=m^2+ k^2-\beta{k}^2(k^2 a^2)^\alpha+\mathcal O(k^6 a^4),
\label{kg5}
\end{eqnarray}
where the third term breaks the Lorentz symmetry.

We turn to consider the energy-momentum relation of Dirac fermions
on a lattice. The Dirac equation in the continuum space-time
\begin{eqnarray}
(i\not\!\partial-m)\psi(x)=0, \label{diraceq}
\end{eqnarray}
and the energy-momentum relation is
\begin{eqnarray}
E= k^2+m^2,\quad -\infty<k<+\infty, \label{diracdr}
\end{eqnarray}
where $k$ is the 3-momentum of fermions. On a spatial lattice
(time continuum), one uses \cite{kogut}
\begin{eqnarray}
\not\!\partial\psi(x)=\gamma_\mu\partial^\mu\psi(x)\Rightarrow\gamma_\mu\frac{[\psi(x+a^\mu)-\psi(x-a^\mu)]}{2a}.
\label{diraceq1}
\end{eqnarray}
where $a^\mu\equiv an^\mu$ and
\begin{eqnarray}
\psi(x+a^\mu)=\psi(k)e^{ik_\mu an^\mu}. \label{diraceq01}
\end{eqnarray}
As a result, the energy-momentum relation of fermion fields on a
lattice is
\begin{eqnarray}
E=\pm \frac{\sin(ka)}{a}.
\label{ddis-rel}
\end{eqnarray}
For $k\cdot a\ll 1$ the energy-momentum relation becomes
\begin{eqnarray}
E=\pm k+ \mathcal O(k^3a^2), \label{ddis-rel2}
\end{eqnarray}
the usual energy-momentum relation. However, Eq.~(\ref{ddis-rel})
has a problem of fermion doubling as shown in Fig.~1, $ka=\pm \pi$
also present fermion spices.

\begin{figure}[t]
 \center
 \includegraphics[width=0.5\columnwidth]{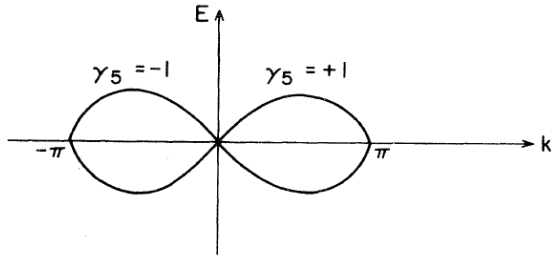}\\
\caption{
Spectrum of the native lattice Dirac equation. } \label{vexf}
\end{figure}
A chiral symmetry breaking term is necessarily added into
Hamiltonian $\mathcal H$ so that $E$ {\it vs} $k$ dose not have
secondary minima at $ka=\pm \pi$. Considering two-component
fermions on a spatial lattice with Wilson term \cite{wilson1}, the
new Hamiltonian is
\begin{eqnarray}
H=&-&\frac{i}{2a}\sum_n \psi^\dag(n)\alpha[\psi(n+1)-\psi(n-1)]\nonumber\\
&+& m \sum_n \bar{\psi}(n)\psi(n) +
\frac{B}{2a}\sum_n\bar{\psi}(n)[2\psi(n)-\psi(n+1)-\psi(n-1)],
\label{fh}
\end{eqnarray}
The equation of motion for $\psi$ is
\begin{eqnarray}
i\dot\psi(n)=-\frac{i}{2a}\gamma_5[\psi(n+1)-\psi(n-1)]+
\frac{B}{2a}\gamma_0[2\psi(n)-\psi(n+1)-\psi(n-1)].
\label{drf}
\end{eqnarray}
Substituting a plane wave $\psi={\rm exp} (iEt-ikna)$ solution into Eq.~(\ref{drf}), one obtains the energy-momentum relation
\begin{eqnarray}
E^2=m^2+\frac{\sin^2ka}{a^2}+4B^2\frac{\sin^4(ka/2)}{a^2}.
\label{drf3}
\end{eqnarray}
For low-energy particles $ka\rightarrow 0$ Eq.~(\ref{drf3}) reduces to
\begin{eqnarray}
E^2 &\simeq & k^2+m^2+\frac{1}{4}B^2k^4a^2+\mathcal O(k^6a^4),
\label{drf4}
\end{eqnarray}
where the third term ($B$-term) violates the Lorentz symmetry.
In principle $B$ is a free parameter characterizing the
deviation from the Lorentz symmetry.

From gravitational theories, for low-energy particles with $E\ll
\xi M_{pl}$, an energy-momentum relation is parametrized as
\cite{naturpiran}
\begin{eqnarray}
E^2-p^2-m^2&\simeq&\pm E^2(\frac{E}{\xi_n E_{pl}})^n, \label{pdr}
\end{eqnarray}
where $\xi_2\gtrsim 10^{-9}$ determined by the flaring AGN \cite{piranref7} for photons $m=0$.
Comparing Eq.~(\ref{drf4}) with Eq.~(\ref{pdr}), one
finds that the lattice spacing $a\lesssim 10^{9}/M_{pl}$, which indicates the Lorentz symmetry breaking scale. We will adopt this scale to study effects of the Lorentz symmetry breaking on hight-energy neutrino oscillations.

\section{Neutrino oscillations due to Lorentz invariance violation}\label{noliv}

In this section we study neutrino oscillations due to the Lorentz
symmetry breaking $B$-term in Eq.~(\ref{drf3}). Flavor neutrinos
$(\nu_e,\nu_\mu, \nu_\tau)$ are always produced
and detected 
via their interacting with
intermediate gauge bosons $W^{(\pm)}_\mu$ and $Z_\mu^0$ in the SM.
Due to the parity violation, flavor neutrinos are not the eigenstates
of the Hamiltonian and in principle they are superpositions of the
Hamiltonian eigenstates $|\nu_i\rangle$
\begin{eqnarray}
\mathcal{H}|\nu_i\rangle&=&E_i|\nu_i\rangle,\quad i=1,2,3,
\label{h0}
\end{eqnarray}
where $E_i$ are the energy eigenvalues of the type-$i$ neutrino.
Using Eq.~(\ref{drf4}) ultra-relativistic neutrinos,
the energy-momentum relation can be approximately written as
\begin{eqnarray}
E_i&\approx& k_i+\frac{m^2_i}{2k_i} + \frac{1}{8}B_i^2k_i^3a^2
+..., \label{ei}
\end{eqnarray}
for the type-$i$ neutrino.

Flavor eigenstates and Hamiltonian eigenstates (mass
eigenstates) are related by an unitary transformation represented
by a matrix $U$,
\begin{eqnarray}
|\nu_l\rangle&=&\sum\limits_{i=1}^{3}U_{li}|\nu_i\rangle
\label{ij},
\end{eqnarray}
where the flavor index $l=e,\mu,\tau$. This shows that flavor
eigenstate is a mixing of the Hamiltonian mass eigenstates
$|\nu_i\rangle, (i=1,2,3)$ and {\it vice versa}. Time evolution of flavor
neutrino states is given by
\begin{eqnarray}
|\nu_l(t)\rangle&=&e^{-i\mathcal{H}t}|\nu_l\rangle=\sum\limits^{3}_{i=1}e^{-iE_it}U_{li}|\nu_i\rangle\,,
\label{newbase0}
\end{eqnarray}
indicating, after some time $t$, the evolution of these flavor neutrino
states leads to flavor neutrino oscillations. The probability of
such neutrino oscillations is given by
\begin{eqnarray}
P_{\nu_l\rightarrow\nu_{l'}}=|\langle
\nu_{l'}|\nu_{l}\rangle|^2=\sum\limits_{i,j}|U_{li}U^*_{l'i}U^*_{lj}U_{l'j}|
\cos[(E_i-E_j)t+\varphi_{ll'}] \label{prob},
\end{eqnarray}
where
\begin{eqnarray}
(E_i-E_j)=\frac{(m^2_i-m^2_j)}{2E}-\frac{(B^2_i-B^2_j)}{8}k^3a^2,
\label{edij}
\end{eqnarray}
 and $\varphi_{ll'}={\rm
arg}(U_{li}U^*_{l'i}U^*_{lj}U_{l'j})$
\cite{mohapatra}-\cite{zober}. In the right-handed side of
Eq.~(\ref{edij}), the first term is normal one and the second term
is due to the Lorentz symmetry breaking $B$-term in
Eq.~(\ref{ei}).

For a two-level system of electron and muon neutrinos
($\nu_e,\nu_{\mu}$). The unitary matrix $U$ is explicitly
given by
\begin{eqnarray}
U=\left(%
\begin{array}{cc}
  \cos\theta & \sin\theta \\
  -\sin\theta & \cos\theta \\
\end{array}%
\right), \label{uni1}
\end{eqnarray}
where $\theta$ is a mixing angle.
Eq.~(\ref{ij}) becomes
\begin{eqnarray}
|\nu_e\rangle&=&\cos\theta|\nu_1\rangle+\sin\theta|\nu_2\rangle\,,\nonumber\\
|\nu_\mu\rangle&=&-\sin\theta|\nu_1\rangle+\cos\theta|\nu_2\rangle\,.
\label{newbase}
\end{eqnarray}
The Hamiltonian (\ref{h0}) in the base of the mass eigenstates $|\nu_i\rangle$ is
\begin{eqnarray}
\mathcal H_{\rm mass}&=&\left(%
\begin{array}{cc}
  E_1 & 0 \\
  0 & E_2 \\
\end{array}%
\right)\simeq E+\left(%
\begin{array}{cc}
  m^2_1/2E & 0 \\
  0 & m^2_2/2E \\
\end{array}%
\right)+\frac{1}{16}\left(%
\begin{array}{cc}
  B^2_1a^2E^3 & 0 \\
  0 & B^2_2a^2E^3 \\
\end{array}%
\right) \label{hmass},
\end{eqnarray}
where the leading contribution to the neutrino energy $E_i$ is
obtained by assuming $p_1\approx p_2=E$. By using
Eqs.~(\ref{uni1},\ref{hmass}) the Hamiltonian in the base of
flavor eigenstates is given by
\begin{eqnarray}
\hat{\mathcal H} &=& U{\mathcal H}_{\rm mass}U^\dag\nonumber\\
&=&E+\frac{m_1^2+m_2^2}{4E}+E^3a^2\left(\frac{B_1^2+B_2^2}{16}\right)+\left(\frac{\Delta
m^2_{12}}{4E}+E^3a^2\frac{\Delta B^2_{12}}{16}\right)
\left(%
\begin{array}{cc}
  -\cos2\theta & \sin2\theta \\
  \sin2\theta & \cos2\theta \\
\end{array}%
\right), \label{h-vac}
\end{eqnarray}
where $\Delta B^2_{12}=B^2_2-B^2_1$, $\Delta B^2_{12}=B^2_2-B^2_1$ , $(B_2>B_1)$
and the mixing angle $\theta$ is given by
\begin{eqnarray}
\tan 2\theta&=&\frac{2\hat{\mathcal H}_{12}}{\hat{\mathcal
H}_{22}-\hat{\mathcal H}_{11}}. \label{theta}
\end{eqnarray}

\section{The conversion probability}

Based on Eq.~(\ref{prob}) for the system of two neutrino flavors, the conversion
and the survival probabilities of a particular flavor of neutrino
with the mixing angle $\theta$, can be written as
\begin{eqnarray}
P_{conv}(t,t_i)&=&\sin^22\theta\sin^2(\frac{\Phi}{2}),\label{conv}\\
P_{surv}&=&1-P_{conv}. \label{surv}
\end{eqnarray}
where the oscillation phase $\Phi$ is given by \cite{lunardini}
\begin{eqnarray}
\Phi=\int^t_{t_i}\varepsilon(\tau)d\tau, \label{phase}
\end{eqnarray}
where $t_i$ and $t$ are respectively the initial and final time of
the evolution of the system. In the case for vacuum oscillations,
$\varepsilon$ equals to \cite{mohapatra}
\begin{eqnarray}
\varepsilon=\varepsilon_{12}\equiv\frac{\Delta m_{12}^2}{2E}.\label{ed12}
\end{eqnarray}
In the case that the Lorentz violation is present
in Eq.~(\ref{h-vac}), $\varepsilon$ can be written as
\begin{eqnarray}
\varepsilon =\frac{\Delta m_{12}^2}{2E}+\frac{1}{8}E^3a^2\Delta
B^2_{12}. \label{lveps}
\end{eqnarray}
which shows  $\Delta B^2$ can also generate
neutrino oscillations.
The discussions and calculations are also applied for other two-level
systems of neutrinos,
($\theta_{23},\Delta B^2_{23}$) and ($\theta_{13},\Delta B^2_{13}$).

Using the scaling relation
\begin{eqnarray}
E=E_0(t_0/t)^{2/3}=E_0(1+z),\label{rescal}
\end{eqnarray}
where $t_0\sim 10^{18}s$ is the present epoch, the redshift
$z\equiv(t_0/t)^{2/3}-1$ and $E_0$ is the energy at the presence
epoch, $z\equiv 0$ \cite{lunardini}. We separate the oscillation phase (\ref{phase})
into two parts:
\begin{eqnarray}
\Phi=\Phi_{vac}+\Phi_{LV}.\label{phase2}
\end{eqnarray}
then using Eqs.~(\ref{phase},\ref{lveps}) and (\ref{rescal}), we obtain
the vacuum and LV phases
\begin{eqnarray}
\Phi_{vac}(x,x_i)&=&\frac{3}{10}\frac{\Delta
m^2t_0}{E_0}(x^{\frac{5}{3}}-x_i^{\frac{5}{3}}),
\label{vacx}\\
\Phi_{LV}(x,x_i)&=&\frac{a^2}{8}\Delta
B^2E_0^3t_0(\frac{1}{x_i}-\frac{1}{x}),\label{lvx}
\end{eqnarray}
where $x\equiv t/t_0$ and $x_i\equiv t_i/t_0$. From (\ref{vacx})
and (\ref{lvx}), we find that the neutrino vacuum oscillation does
not occur, $\Delta m^2\rightarrow 0$, however neutrino
oscillations due to the Lorentz symmetry violation take place.

Taking $x_i=0.125$, corresponding to the initial time of neutrino
productions at redshift $z\simeq 3$, and $x=1$, we obtain
\begin{eqnarray}
\Phi_{vac}(1,0.125)&\simeq & \frac{3}{10}\frac{\Delta
m^2}{E_0}t_0,\label{vacx1}\\
\Phi_{LV}(1,0.125)&\simeq &\frac{7}{8}a^2\Delta B^2E_0^3t_0.\label{lvx1}
\end{eqnarray}
Eqs.~(\ref{vacx1}) and (\ref{lvx1}) show that for very high energy
neutrinos, the LV oscillation phase becomes more important than
the vacuum oscillation phase.

\begin{figure}[t]
 \center
 \includegraphics[width=0.4\columnwidth]{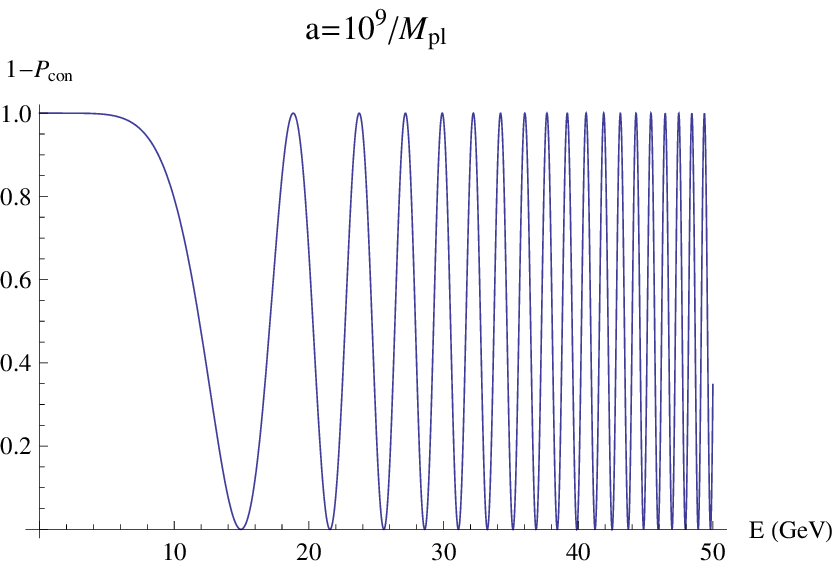}
 \,\,\,\,\,\,\,\includegraphics[width=0.4\columnwidth]{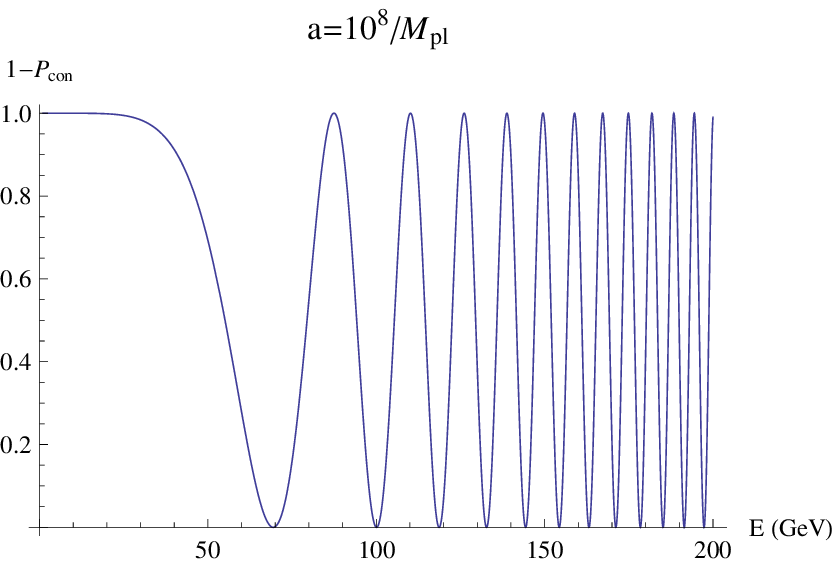}\\
 \includegraphics[width=0.4\columnwidth]{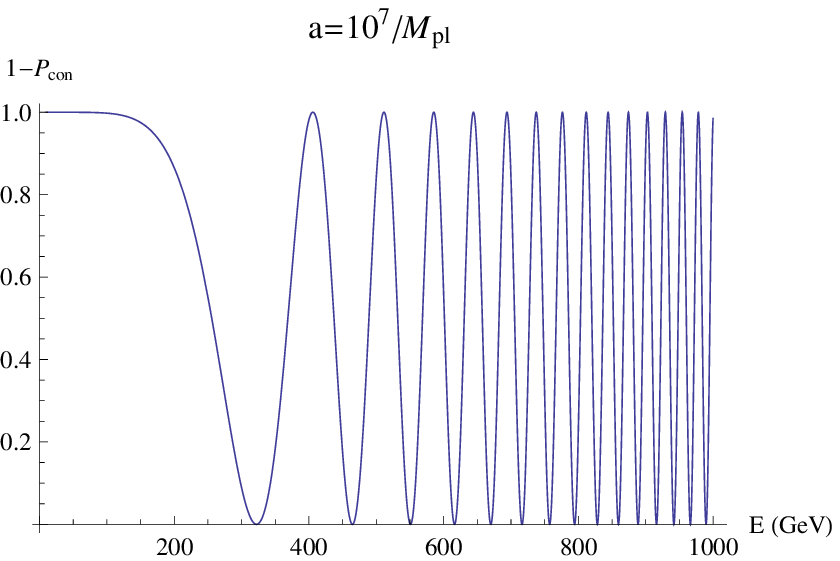}
 \,\,\,\,\,\,\,\includegraphics[width=0.4\columnwidth]{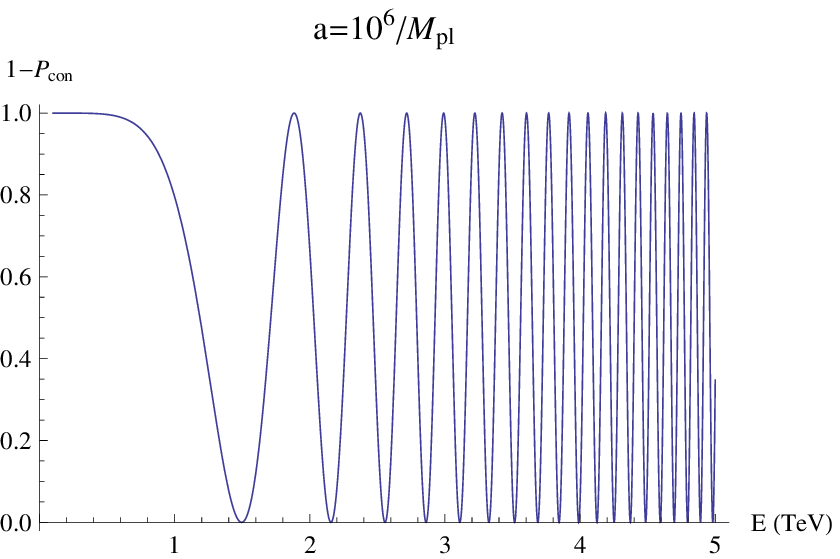}\\
 \includegraphics[width=0.4\columnwidth]{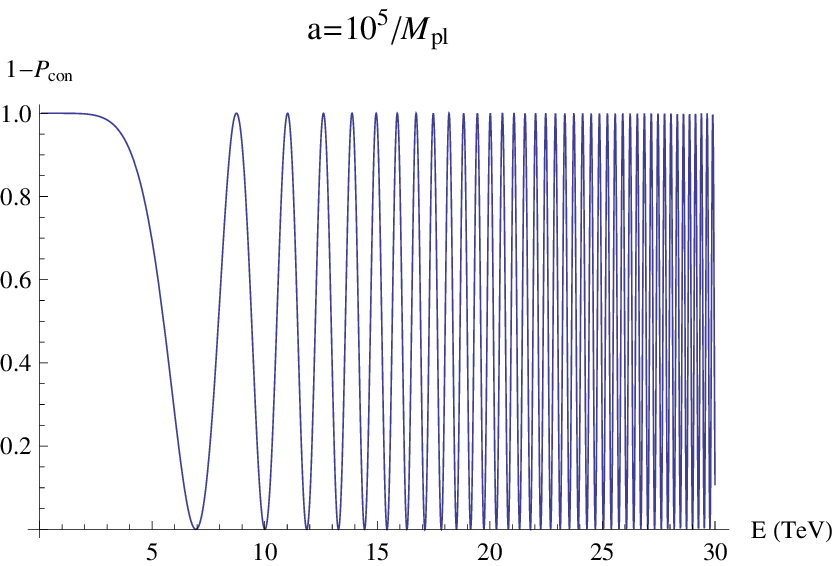}
\caption{
The survival probability
$1-P_{conv}(\nu_\alpha\rightarrow\nu_\beta)$ is plotted as a
function of neutrino energy $E_0$, for different values of the
Lorentz symmetry breaking scale $a$. $\Delta m^2 \approx
10^{-7}eV^2$ and the mixing angle $\sin^22\theta\simeq
1$\,\cite{lunardini} and $|\Delta B^2|  \approx |\Delta m^2|$}.
\label{fig2}
\end{figure}

Since neutrino detectors have a finite accuracy in the
reconstruction of the neutrino energy, by averaging
Eq.~(\ref{conv}) over the interval $\Delta E_0\simeq E_0$, one
computes the conversion probability \cite{lunardini}
\begin{eqnarray}
P_{conv}(E_0)=\frac{1}{\Delta E_0}\int^{3E_0/2}_{E_0/2}dE'P(E').
\label{avp}
\end{eqnarray}
Considering very high energy neutrinos, which are produce at
$z=3$, using Eqs.~(\ref{surv},\ref{vacx1}-\ref{avp}), we plot in
Fig.~(\ref{fig2}) the survival probability as a function of energy
$E_0$. We find that for large neutrino energies, the vacuum
oscillation phase (\ref{vacx1}) is suppressed and its contribution
to the conversion probability
$P_{conv}(\nu_\alpha\rightarrow\nu_\beta)$ is almost zero, and the
conversion probability $P_{conv}(\nu_\alpha\rightarrow\nu_\beta)$
is mainly contributed from the neutrino oscillation phase
(\ref{lvx1}) due to the Lorentz symmetry breaking. This implies
that any observation of high-energy neutrino oscillations
indicates the Lorentz symmetry breaking. In addition, the neutrino
oscillation pattern ($E_0$-dependence) due to the Lorentz symmetry
breaking is very different from the neutrino oscillation pattern
in vacuum. This might provide the possibility that using
high-energy cosmic neutrinos, one can study neutrino oscillation
pattern to gain some insight into the Lorentz symmetry breaking,
in connection with the study of arrival time delay of high-energy
cosmic gamma ray due to the Lorentz symmetry breaking
\cite{naturpiran}. In addition, from the theoretical point view,
it would be interesting to see how the Lorentz violation term
(\ref{drf}-\ref{drf4}) relate to Lorentz violation operators in
effective field theories, see for example Ref.~\cite{nbq}.

\comment{ \rr{One can use these computations in other high energy
neutrino oscillation scenario, such as SME (see \cite{nbq} and
reference therein )}}

\end{document}